\documentclass[preprint,superscriptaddress,showpacs,preprintnumbers,amsmath,amssymb,eqsecnum]{revtex4}

\usepackage{graphicx}
\usepackage{dcolumn}
\usepackage{bm}

\def\lesssim{\ \raise.3ex\hbox{$<$}\kern-0.8em\lower.7ex\hbox{$\sim$}\ }
\def\gesim{\ \raise.3ex\hbox{$>$}\kern-0.8em\lower.7ex\hbox{$\sim$}\ }

\begin{document}

\preprint{APS/123-QED}

\title{Supercurrent behavior of low-energy Bogoliubov phonon and anomalous tunneling effect in a Bose-Einstein condensate}
\author{Yoji Ohashi}%
\email{yohashi@rk.phys.keio.ac.jp}
\author{Shunji Tsuchiya}
\email{tsuchiya@rk.phys.keio.ac.jp}

\affiliation{Department of Physics, Keio University, 3-14-1 Hiyoshi, Kohoku-ku, Yokohama 223-8522, Japan.}
\affiliation{CREST(JST), 4-1-8 Honcho, Saitama 332-0012, Japan}
\date{\today}


\begin{abstract}
We investigate tunneling properties of Bogoliubov mode in a Bose-Einstein condensate. Using an exactly solvable model with a $\delta$-functional barrier, we show that each component in the two-component wavefunction $(u,v)$ of low-energy Bogoliubov phonon has the same form as the condensate wavefunction in the supercurrent state. As a result, the currents $J_u$ and $J_v$ associated with $u$ and $v$, respectively, have the same tunneling properties as those of supercurrent carried by condensate. Thus, the tunneling of low-energy Bogoliubov phonon described by the tunneling of these two currents shows perfect transmission. We also show that the supercurrent behaviors of Bogoliubov phonon still exist in the presence of supercurrent carried by condensate, except in the critical supercurrent state. In the critical current state, the perfect transmission is absent, because $J_u$ or $J_v$ exceeds their upper limit given by the critical value of the supercurrent associated with the condensate. Our results consistently explain the recently proposed two tunneling phenomena associated with Bogoliubov phonon, namely, the anomalous tunneling effect (perfect transmission in the low-energy limit) and the breakdown of the perfect transmission in the critical supercurrent state.
\end{abstract}

\maketitle


\section{\label{sec1}Introduction}
Recently, novel tunneling phenomena of Bogoliubov phonon have been theoretically predicted in superfluid Bose gases. Kovrizhin and co-workers\cite{Kov2,Kovrizhin,Kagan} clarified the perfect transmission of low-energy Bogoliubov phonon across a potential barrier, which is referred to as the anomalous tunneling effect. Danshita and co-workers\cite{Danshita} showed that the anomalous tunneling effect also occurs in the supercurrent state, as far as the magnitude of the supercurrent is less than the critical current\cite{Danshita}. In the critical supercurrent state, the perfect transmission is not obtained\cite{Danshita}, irrespective of the relative direction between the momentum of Bogoliubov phonon and superflow. They also extended their work to a Bose condensate in an optical lattice\cite{Danshita2,Danshita3}. Although the anomalous tunneling effect has not been observed yet, a cold atom gas may be useful to examine this interesting tunneling phenomenon. With this regard, we briefly note that a double-well trap has been realized in cold atom physics\cite{Andrews}, and a kind of Josephson effect has been observed\cite{Albiez}. 
\par
For the mechanism of the anomalous tunneling effect, various key issues have been discussed, such as resonance tunneling\cite{Kagan}, localized state near the potential barrier\cite{Danshita}, similarity between the wavefunction of Bogoliubov mode and condensate wavefunction in the low energy limit\cite{Kato}, and coupling of quasiparticle current with supercurrent near the barrier\cite{Tsuchiya}. However, despite these great efforts, no consistent explanation for the anomalous tunneling and the breakdown of this effect in the critical supercurrent state has not been given yet. Since the appearance of Bogoliubov phonon is one of the most fundamental phenomena in the superfluid phase\cite{Bogoliubov}, clarifying physical properties of this collective mode is a very important issue in the research of superfluidity. 
\par
In this paper, we investigate tunneling properties of Bogoliubov excitations in a weakly interacting Bose superfluid at $T=0$. We treat the condensate wavefunction and Bogoliubov excitations within the Gross Pitaevskii (GP) equation and Bogoliubov equations, respectively. Using an exactly solvable model, we analytically show that the tunneling mechanism of low-energy Bogoliubov phonon is the same as that of ordinary supercurrent associated with the condensate. Since the supercurrent is well known to tunnel through a barrier without reflection, our result immediately explains the anomalous tunneling effect associated with Bogoliubov phonon.  However, in contrast to the ordinary supercurrent, the Bogoliubov phonon consists of two current components ($\equiv J_u$ and $J_v$), whose directions are opposite to each other. These counterpropagating currents have been recently observed experimentally by using Bragg spectroscopy\cite{Bragg}. In this paper, we show that they have the same upper limit which equals the upper limit of the ordinary supercurrent (critical supercurrent $J_c$). In the critical supercurrent state, $J_u$ or $J_v$ is shown to always exceed $J_c$, so that the supercurrent behavior of Bogoliubov mode (perfect transmission) is destroyed. This result gives simple and physical explanation for the breakdown of the anomalous tunneling effect in the critical supercurrent state predicted in Ref. \cite{Danshita}. 
\par
This paper is organized as follows. In Sec. II, we present our tunneling model with a $\delta$-functional potential barrier. In this model, exact solutions for the GP and Bogoliubov equations have been derived in Refs. \cite{Kovrizhin,Danshita}. Since we use these solutions in later sections, we summarize them in this section. For their detailed derivations, we refer to Refs. \cite{Kovrizhin,Danshita}. In Sec. III, we consider the case in the absence of the supercurrent. Here, we show that the anomalous tunneling effect can be explained as a result of the supercurrent behavior of Bogoliubov mode. This result is extended to the supercurrent state in Sec. IV. In Sec. V, we discuss how the breakdown of the anomalous tunneling effect in the critical supercurrent state can be understood based on our results obtained in Secs. III and IV. Throughout this paper, we set $\hbar=1$. 
\par

\begin{figure}
\centerline{\includegraphics[width=10cm]{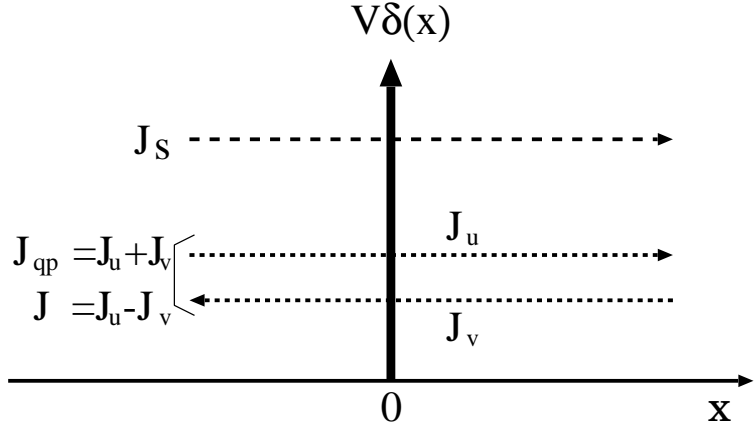}}
\caption{Schematic picture of our model. A Bose superfluid is separated by the barrier potential $V\delta(x)$, and we consider the tunneling of Bogoliubov phonon injected from $x=-\infty$. We examine both the cases with and without supercurrent $J_s$ ($>0$) carried by condensate. The quasiparticle current $J_{qp}$ carried by Bogoliubov phonon is described by the sum of two currents as $J_{qp}=J_u+J_v$, while the probability current density $J$ is given by $J=J_u-J_v$, where $J_u$ and $J_v$ are defined by Eq. (\ref{eq.22}). We note that the flow direction of $J_v$ is opposite to that of $J_u$.}
\label{fig1}
\end{figure}

\section{Model tunneling problem, condensate wavefunction and Bogoliubov excitations}
\par
We consider a superfluid Bose gas at $T=0$, which is separated by the potential barrier $V(x)=V\delta(x)$ (Fig. \ref{fig1}). Since the system is uniform in the $y$- and $z$-direction, our model is essentially a one-dimensional system. Thus, we only retain the $x$-direction, and ignore the $y$- and $z$-direction. We also ignore effects of a harmonic trap potential, for simplicity. The latter simplification is allowed when a box-shaped trap is considered\cite{Meyrath}. 
\par
In this model, exact solutions for the GP equation and Bogoliubov equations have been derived in Refs. \cite{Kovrizhin,Danshita}. Since we use these solutions in later sections, we summarize them in this section. For detailed derivations of the exact solutions, we refer to Refs. \cite{Kovrizhin,Danshita}.
\par
The Gross-Pitaevskii (GP) equation for the condensate wavefunction $\Psi(x)$ is given by\cite{Pitaevskii}
\begin{eqnarray}
\Bigl(
-{1 \over 2m}{d^2 \over dx^2}-\mu+V\delta(x)+g|\Psi(x)|^2
\Bigr)\Psi(x)=0.
\label{eq.1}
\end{eqnarray}
Here, $m$ is the mass of a Bose atom, and $\mu$ is the chemical potential. $g$ is a repulsive interaction between Bose atoms. Introducing the scaled variables,
\begin{eqnarray}
{\bar \Psi}({\bar x})\equiv {\Psi(x) \over \sqrt{n_0}},~~{\bar \mu}\equiv {\mu \over gn_0},~~{\bar V}\equiv{V \over gn_0\xi},~~{\bar x}\equiv{x \over \xi},
\label{eq.2}
\end{eqnarray}
we can write Eq. (\ref{eq.1}) in the dimensionless form
\begin{eqnarray}
\Bigl(
-{1 \over 2}{d^2 \over d{\bar x}^2}-{\bar \mu}+{\bar V}\delta({\bar x})+|{\bar \Psi({\bar x})}|^2
\Bigr){\bar \Psi}({\bar x})=0.
\label{eq.3}
\end{eqnarray}
In Eq. (\ref{eq.2}), $n_0\equiv|\Psi(x=\pm\infty)|^2$ is the condensate density far away from the barrier, and $\xi\equiv 1/\sqrt{mgn_0}$ is the healing length.  The scaled chemical potential ${\bar \mu}$ equals unity in the absence of supercurrent. In the supercurrent state with momentum $q$, one finds $\mu=1+q^2/2$\cite{Danshita}. In the following, we simply write $({\bar \Psi},{\bar \mu},{\bar V},{\bar x})$ as $(\Psi,\mu,V,x)$. 
\par
The boundary conditions at $x=0$ are given by
\begin{eqnarray}
\Psi(+0)=\Psi(-0),
\label{eq.4}
\end{eqnarray}
\begin{eqnarray}
\Bigl({d\Psi \over dx}\Bigr)_{x=+0}
-
\Bigl({d\Psi \over dx}\Bigr)_{x=-0}
=2V\Psi(0).
\label{eq.5}
\end{eqnarray}
We solve the GP equation (\ref{eq.3}) together with the boundary conditions in Eqs. (\ref{eq.4}) and (\ref{eq.5}). The condensate wavefunction $\Psi_q(x)$ in the supercurrent state is obtained as\cite{Danshita}
\begin{equation}
\Psi_q(x)=e^{i[qx-{\rm sgn}(x)\theta_q]}
[\gamma(x)-iq{\rm sgn}(x)],
\label{eq.6}
\end{equation}
where
\begin{equation}
\gamma(x)\equiv\sqrt{1-q^2}\tanh(\sqrt{1-q^2}[|x|+x_0]),
\label{eq.7}
\end{equation}
\begin{equation}
e^{i\theta_q}\equiv
{\gamma(0)-iq \over \sqrt{\gamma(0)^2+q^2}}.
\label{eq.8}
\end{equation}
$x_0$ in Eq. (\ref{eq.7}) is determined by
\begin{equation}
\gamma(0)^3+V\Bigl(\gamma(0)^2+q^2\Bigr)-\Bigl(1-q^2\Bigr)\gamma(0)=0.
\label{eq.9}
\end{equation}
Equation (\ref{eq.9}) is obtained from the boundary condition in Eq. (\ref{eq.5}).
\par
We note that the magnitude of the condensate wavefunction $\Psi_q(x)$ in Eq. (\ref{eq.6}) is suppressed near the barrier. However, the supercurrent density $J_s$ is independent of $x$ as,
\begin{equation}
J_s={\rm Im}\Bigl[\Psi_q^*(x){d \over dx}\Psi_q(x)\Bigr]=q. 
\label{eq.10}
\end{equation}
Namely, the supercurrent is conserved in the whole system.
\par
For a given condensate wavefunction $\Psi_q(x)$, the two-component wavefunction $(u,v)$ of the Bogoliubov mode is obtained from the Bogoliubov equations\cite{Pitaevskii}. Using the scaled variables in Eq. (\ref{eq.2}), one can write the Bogoliubov equations in the dimensionless forms
\begin{equation}
\Bigl(
-{1 \over 2}{d^2 \over d{\bar x}^2}-{\bar \mu}+{\bar V}\delta({\bar x})+2|{\bar \Psi}_q({\bar x})|^2
\Bigr){\bar u}({\bar x})
-{\bar \Psi}_q({\bar x})^2{\bar v}({\bar x})={\bar E}{\bar u}({\bar x}),
\label{eq.11}
\end{equation}
\begin{equation}
\Bigl(
-{1 \over 2}{d^2 \over d{\bar x}^2}-{\bar \mu}+{\bar V}\delta({\bar x})+2|{\bar \Psi}_q({\bar x})|^2
\Bigr){\bar v}({\bar x})
-{\bar \Psi}_q^*({\bar x})^2{\bar u}({\bar x})=-{\bar E}{\bar u}({\bar x}).
\label{eq.12}
\end{equation}
Here, the Bogoliubov wavefunction $(u(x),v(x))$ and the energy $E$ have been scaled as $({\bar u}({\bar x}),{\bar v}({\bar x}))\equiv (u(x),v(x))/\sqrt{n_0}$ and ${\bar E}\equiv E/gn_0$, respectively. In the following, we omit the bars in the Bogoliubov equations (\ref{eq.11}) and (\ref{eq.12}). The boundary conditions for the Bogoliubov mode $(u,v)$ at $x=0$ are given by
\begin{equation}
u(+0)=u(-0),
~~~
\Bigl({du(x) \over dx}\Bigr)_{x=+0}-\Bigl({du(x) \over dx}\Bigr)_{x=-0}=2Vu(0),
\label{eq.13}
\end{equation}
\begin{equation}
v(+0)=v(-0),
~~~
\Bigl({dv(x) \over dx}\Bigr)_{x=+0}-\Bigl({dv(x) \over dx}\Bigr)_{x=-0}=2Vv(0).\label{eq.14}
\end{equation}
\par
To construct $(u,v)$ satisfying the boundary conditions in Eqs. (\ref{eq.13}) and (\ref{eq.14}), we need particular solutions of the Bogoliubov equations (\ref{eq.11}) and (\ref{eq.12}) for $x\ge 0$ and $x\le 0$. For a given condensate wavefunction $\Psi_q(x)$ and energy $E$, the coupled equations (\ref{eq.11}) and (\ref{eq.12}) have four particular solutions $(u_n,v_n)$ $(n=1,2,3,4)$, given by\cite{Kovrizhin,Danshita}
\begin{eqnarray}
\displaystyle
\left\{
\begin{array}{l}
\displaystyle
u_n=e^{i[(p_n+q)x-{\rm sgn}(x)\theta_q]}
\Bigl[
\Bigl(1+{p_n^2 \over 2E}\Bigr)\gamma(x)
-i{\rm sgn}(x)
\Bigl(
q+{p_n \over 2E}(1-q^2-\gamma(x)^2+E)
+{p_n^3 \over 4E}
\Bigr)
\Bigr],
\\
\displaystyle
v_n=e^{i[(p_n-q)x+{\rm sgn}(x)\theta_q]}
\Bigl[
\Bigl(1-{p_n^2 \over 2E}\Bigr)\gamma(x)
+i{\rm sgn}(x)
\Bigl(
q+{p_n \over 2E}(1-q^2-\gamma(x)^2-E)
+{p_n^3 \over 4E}
\Bigr)
\Bigr].
\end{array}
\right.
\nonumber
\\
\label{eq.16}
\end{eqnarray}
(We note that Eq. (\ref{eq.16}) is not normalized.) The momenta $p_n$ ($n=1,2,3,4$) are obtained from the expression for the Bogoliubov excitation spectrum\cite{Danshita}
\begin{equation}
E=p_nq+\sqrt{{p_n^2 \over 2}\Bigl({p_n^2 \over 2}+2\Bigr)}.
\label{eq.17}
\end{equation}
Among the four solutions, two of them ($n=1$ and 2) describe propagating wave characterized by real momenta ($p_1$ and $p_2$). The remaining two solutions ($n=3$ and 4) describe localized states having complex momenta ($p_3$ and $p_4$). While only the propagating solutions are necessary in considering a uniform system, one has to also take into account the localized solutions in the present inhomogeneous system. Indeed, in Sec. III, we show that the localized states appear near the barrier. 
\par
In the low energy region ($E\ll 1$), $p_n$ ($n=1,2,3,4$) reduce to\cite{note}, within the accuracy of $O(E)$,
\begin{eqnarray}
\left\{
\begin{array}{l}
\displaystyle
p_1={E \over 1+q},\\
\displaystyle
p_2=-{E \over 1-q},\\
\displaystyle
p_3=2i\sqrt{1-q^2}+{qE \over 1-q^2},\\
\displaystyle
p_4=-2i\sqrt{1-q^2}+{qE \over 1-q^2}.
\end{array}
\label{eq.17b}
\right.
\end{eqnarray}
\par
Using these particular solutions, we construct the Bogoliubov wavefunction $(u,v)$. Assuming that the incident Bogoliubov phonon comes from $x=-\infty$, we set
\begin{eqnarray}
\left(
\begin{array}{l}
u(x)\\v(x)
\end{array}
\right)
=
\left(
\begin{array}{l}
u_<(x)\\v_<(x)
\end{array}
\right)
\theta(-x)+
\left(
\begin{array}{l}
u_>(x)\\v_>(x)
\end{array}
\right)
\theta(x),
\label{eq.16b}
\end{eqnarray}
Here, $\theta(x)$ is the step function, and
\begin{eqnarray}
\left(
\begin{array}{l}
u_<(x)\\v_<(x)
\end{array}
\right)
=
\left(
\begin{array}{l}
u_1(x)\\v_1(x)
\end{array}
\right)+
A
\left(
\begin{array}{l}
u_2(x)\\v_2(x)
\end{array}
\right)
+
B\left(
\begin{array}{l}
u_4(x)\\v_4(x)
\end{array}
\right)~~~~~(x\le 0),
\label{eq.18}
\end{eqnarray}
\begin{eqnarray}
\left(
\begin{array}{l}
u_>(x)\\v_>(x)
\end{array}
\right)
=
C
\left(
\begin{array}{l}
u_1(x)\\v_1(x)
\end{array}
\right)
+
D\left(
\begin{array}{l}
u_3(x)\\v_3(x)
\end{array}
\right)~~~~~~~~~~~~~~~(x\ge0).
\label{eq.19}
\end{eqnarray}
The coefficients $(A,B,C,D)$ are determined so that the boundary conditions in Eqs. (\ref{eq.13}) and (\ref{eq.14}) can be satisfied. We will give their detailed expressions in later sections.
\par
Once $(A,B,C,D)$ are determined, the transmission probability is obtained from the ratio of the probability current density $J\equiv J_u-J_v$ for the incident wave to that for the transmitted wave. Here, $J_u$ and $J_v$ are given by\cite{note9} 
\begin{eqnarray}
\left\{
\begin{array}{l}
\displaystyle
J_u={\rm Im}\Bigl[u(x)^*{d \over dx}u(x)\Bigr],\\
\displaystyle
J_v=-{\rm Im}\Bigl[v(x)^*{d \over dx}v(x)\Bigr].
\end{array}
\right.
\label{eq.20}
\end{eqnarray}
We note that the quasiparticle current density $J_{\rm qp}$ carried by Bogoliubov phonon has the different form from the probability current density as $J_{\rm qp}=J_u+J_v$. To see the difference between $J$ and $J_{\rm qp}$, it is useful to consider a uniform system, in which the Bogoliubov equations give the plane wave solution\cite{Pitaevskii}
\begin{eqnarray}
\left\{
\begin{array}{l}
\displaystyle
u_p=\sqrt{{1 \over 2}\Bigl({p^2/2+1 \over E}+1\Bigr)}e^{ipx},\\
\displaystyle
v_p=\sqrt{{1 \over 2}\Bigl({p^2/2+1 \over E}-1\Bigr)}e^{ipx}.
\end{array}
\right.
\label{eq.21}
\end{eqnarray}
Here, $E=\sqrt{(p^2/2)(p^2/2+2)}$ is the Bogoliubov excitation spectrum. Substituting Eq. (\ref{eq.21}) into Eq. (\ref{eq.20}), we obtain ($p>0$),
\begin{eqnarray}
\left\{
\begin{array}{l}
\displaystyle
J_u={1 \over 2}\Bigl({p^2/2+1 \over E}+1\Bigr)p~~~(>0),\\
\displaystyle
J_v=-{1 \over 2}\Bigl({p^2/2+1 \over E}-1\Bigr)p~~~(<0).
\end{array}
\right.
\label{eq.22}
\end{eqnarray}
Equation (\ref{eq.22}) shows that the leading terms of $J_u$ and $J_v$ with respect to $p$ are constant in the low energy limit. While they dominantly contribute to the probability current density $J$, they are cancelled out in the quasiparticle current $J_{\rm qp}$, because the flow directions of $J_u$ and $J_v$ are opposite to each other. The contribution to $J_{\rm qp}$ comes from higher order terms in $J_u$ and $J_v$ in terms of $p$. In Sec. III, we find that the cancellation of the leading terms of $J_u$ and $J_v$ also occur in the presence of the barrier, as schematically shown in Fig. \ref{fig1}. Thus. in considering the tunneling of low-energy Bogoliubov phonon, the tunneling properties of each component $J_u$ and $J_v$ are more crucial than the quasiparticle tunneling current $J_{\rm qp}$ given by the sum of them. In Secs. III and IV, we show that each $J_u$ and $J_v$ has the same tunneling properties as those of supercurrent in the low energy region, leading to the anomalous tunneling of Bogoliubov phonon.
\par
\section{Supercurrent behavior of low-energy Bogoliubov mode}
\par
In this section, we first consider the tunneling of low-energy Bogoliubov phonon in the absence of supercurrent. In this case, the condensate wavefunction in Eq. (\ref{eq.6}) reduces to
\begin{equation}
\Psi_{q=0}(x)=\gamma(x)=\tanh(|x|+x_0).
\label{eq.3.1}
\end{equation}
Since we are interested in the anomalous tunneling effect, we consider the low energy region ($E\ll 1$). In this regime, the Bogoliubov excitation spectrum has the linear dispersion $E=p$, where $p$ ($>0$) is the momentum of incident wave coming from $x=-\infty$. For the momenta of the four particular solutions in Eq. (\ref{eq.17b}), we may take $(p_1,p_2,p_3,p_4)=(p,-p,2i,-2i)$ within the accuracy of $O(p)$. Determining the coefficients $(A,B,C,D)$ in Eqs. (\ref{eq.18}) and (\ref{eq.19}), we obtain\cite{Kovrizhin,Danshita}, to the accuracy of $O(p)$,
\begin{eqnarray}
\left(
\begin{array}{l}
u_<(x)\\v_<(x)
\end{array}
\right)
=
\left(
\begin{array}{l}
u_I(x)\\v_I(x)
\end{array}
\right)e^{ipx}
+
i\alpha p
\left(
\begin{array}{l}
u_R(x)\\v_R(x)
\end{array}
\right)e^{-ipx}
+
\Bigl({i \over 2}+\beta p\Bigr)\left(
\begin{array}{l}
u_L(x)\\v_L(x)
\end{array}
\right),
\label{eq.3.2}
\end{eqnarray}
\begin{eqnarray}
\left(
\begin{array}{l}
u_>(x)\\v_>(x)
\end{array}
\right)
=
(1-i\eta p)
\left(
\begin{array}{l}
u_T(x)\\v_T(x)
\end{array}
\right)e^{ipx}
-
\Bigl({i \over 2}+\beta p\Bigr)\left(
\begin{array}{l}
u_L(x)\\v_L(x)
\end{array}
\right).
\label{eq.3.3}
\end{eqnarray}
The coefficients $(\alpha,~\beta,~\eta)$ are given by
\begin{eqnarray}
\left\{
\begin{array}{l}
\displaystyle
\alpha=-{1 \over 2}(1-\gamma_0)
{2+\gamma_0+\gamma_0^2 \over (1+\gamma_0^2)\gamma_0},\\
\displaystyle
\beta={1 \over 2}\Bigl({1 \over 2}-{1 \over \gamma_0}\Bigr),\\
\displaystyle
\eta=\alpha-{2\gamma_0 \over 1+\gamma_0^2},
\end{array}
\right.
\label{eq.3.4}
\end{eqnarray}
where we have simply written $\gamma_0=\gamma(x=0)$. In Eqs. (\ref{eq.3.2}) and (\ref{eq.3.3}), the incident wave $(u_I,v_I)$ and transmitted wave $(u_T,v_T)$ are obtained from the particular solution $(u_1,v_1)$ in Eq. (\ref{eq.16}), while the reflected wave $(u_R,v_R)$ is obtained from $(u_2,v_2)$. In the low energy limit, they are given by
\begin{eqnarray}
\left(
\begin{array}{c}
u_I \\v_I
\end{array}
\right)
=
\left(
\begin{array}{l}
\displaystyle
\gamma(x)+{i \over 2}[1-\gamma(x)^2]+{p \over 2}[\gamma(x)+i]\\
\displaystyle
\gamma(x)-{i \over 2}[1-\gamma(x)^2]-{p \over 2}[\gamma(x)-i]
\end{array}
\right),
\label{eq.3.5}
\end{eqnarray}
\begin{eqnarray}
\left(
\begin{array}{c}
u_R \\v_R
\end{array}
\right)
=
\left(
\begin{array}{c}
u_T \\v_T
\end{array}
\right)
=
\left(
\begin{array}{l}
\displaystyle
\gamma(x)-{i \over 2}[1-\gamma(x)^2]+{p \over 2}[\gamma(x)-i]\\
\displaystyle
\gamma(x)+{i \over 2}[1-\gamma(x)^2]-{p \over 2}[\gamma(x)+i]
\end{array}
\right).
\label{eq.3.6b}
\end{eqnarray}
The localized component $(u_L,v_L)$ in Eqs. (\ref{eq.3.2}) and (\ref{eq.3.3}) is obtained from $(u_3,v_3)$ and $(u_4,v_4)$, which has the form
\begin{eqnarray}
\left(
\begin{array}{c}
u_L \\v_L
\end{array}
\right)
=
\left(
\begin{array}{l}
\displaystyle
-[1-\gamma(x)^2]+p[1-\gamma(x)]\\
\displaystyle
[1-\gamma(x)^2]+p[1-\gamma(x)]
\end{array}
\right).
\label{eq.3.6c}
\end{eqnarray}
We note that, in the last terms of Eqs. (\ref{eq.3.2}) and (\ref{eq.3.3}), the exponential factors $e^{\pm 2x}$ appearing in $(u_3,v_3)$ and $(u_4,v_4)$ have been absorbed into $(u_L,v_L)$ by using the identity
\begin{equation}
e^{-2(|x|+x_0)}[1+\gamma(x)]=1-\gamma(x).
\label{eq.3.7}
\end{equation}
\par
To show the supercurrent behavior of low-energy Bogoliubov phonon in the tunneling process, we expand Eqs. (\ref{eq.3.2}) and (\ref{eq.3.3}) in terms of $p$ to the first order. For example, $u_>(x)$ then becomes
\begin{eqnarray}
u_>(x)
&=&\gamma(x)+ip{\gamma(x)-\gamma_0 \over \gamma_0}+ipx\gamma(x)
\nonumber
\\
&+&
ip{\gamma_0 \over 1+\gamma_0^2}\gamma(x)
+{p \over 2}{\gamma_0 \over 1+\gamma_0^2}[1-\gamma(x)^2]
+{p \over 2}\gamma(x)+{p \over 2}x[1-\gamma(x)^2].
\label{eq.3.8}
\end{eqnarray}
Using the identity
\begin{equation}
{d \over dx}\gamma(x)={\rm sgn}(x)
\Bigl(1-\gamma(x)^2\Bigr),
\label{eq.3.9}
\end{equation}
we find that Eq. (\ref{eq.3.8}) is equivalent to the expression,
\begin{equation}
u_>(x)=\sqrt{1+p}e^{ip{\gamma_0 \over 1+\gamma_0^2}}e^{ipx}
\Bigl[
\gamma(\sqrt{1+p}x+{p\gamma_0 \over 2(1+\gamma_0^2)})
+ip{\gamma(x)-\gamma_0 \over \gamma_0}
\Bigr],
\label{eq.3.10}
\end{equation}
within the accuracy of $O(p)$. Expanding the condensate wave function in Eq. (\ref{eq.6}) in the long wave length limit, we obtain ($x>0$)
\begin{equation}
\Psi_q(x)=e^{iqx}\Bigl(\gamma(x)+iq{\gamma(x)-\gamma_0 \over \gamma_0}\Bigr).
\label{eq.3.11}
\end{equation}
Comparing Eq. (\ref{eq.3.10}) with (\ref{eq.3.11}), we find that $u_>(x)$ essentially has the same form as the condensate wavefunction $\Psi_{q=p}(x)$ in the {\it supercurrent state} with momentum $p$. The same conclusion is also obtained for $u_<(x)$. Introducing $y_+\equiv\sqrt{1+p}x$ and
\begin{eqnarray}
{\tilde u}(y_+)\equiv {1 \over \sqrt{1+p}}e^{-ip{\gamma_0 \over 1+\gamma_0^2}}u(x=y_+/\sqrt{1+p}),
\label{eq.3.12}
\end{eqnarray}
we obtain
\begin{equation}
{\tilde u}(y_+)={\tilde \Psi}_p(y_+,Y_+).
\label{eq.3.13}
\end{equation}
Here, ${\tilde \Psi}_p$ is given by Eq. (\ref{eq.6}), where $x_0$ in $\gamma$ is replaced by 
\begin{equation}
Y_+= x_0+{\gamma_0 \over 2(1+\gamma_0^2)}p.
\label{eq.3.13b} 
\end{equation}
\par
The same analysis is applicable to $v(x)$. From Eqs. (\ref{eq.3.2})-(\ref{eq.3.6c}), $v_p(x)$ is found to be related to $u_p(x)$ as $v_p(x)=u_{-p}(x)^*$. Using this, we obtain
\begin{eqnarray}
{\tilde v}(y_-)={\tilde \Psi}_p(y_-,Y_-),
\label{eq.3.14}
\end{eqnarray}
where $y_-=\sqrt{1-p}x$, 
\begin{equation}
Y_-=x_0-{\gamma_0 \over 2(1+\gamma_0^2)}p, 
\label{eq.3.14b}
\end{equation}
and 
\begin{eqnarray}
{\tilde v}(y_-)\equiv 
{1 \over \sqrt{1-p}}e^{-ip{\gamma_0 \over 1+\gamma_0^2}}v(x=y_-\sqrt{1-p}).
\label{eq.3.15}
\end{eqnarray}
\par
Equations (\ref{eq.3.13}) and (\ref{eq.3.14}) clearly show that, in the low energy region, $u(x)$ and $v(x)$ have the same properties as the condensate wavefunction in the supercurrent state. Namely, the currents $J_u$ and $J_v$ tunnel through the barrier without reflection in the low energy region. The tunneling of low energy Bogoliubov phonon described by the wavefunction $(u,v)$ is also not accompanied by reflection, which naturally explains the anomalous tunneling effect\cite{Kovrizhin}. The coefficient of the reflected wave component given by the second term in the right hand side of Eq. (\ref{eq.3.2}) is proportional to $p$, so that the reflection probability is proportional to $p^2$. Namely, deviation from the perfect transmission starts from $O(p^2)$. 
\par
Since $u(x)$ and $v(x)$ have the same form as the condensate wavefunction, the currents $J_u$ and $J_v$ are immediately obtained from Eq. (\ref{eq.10}) as 
\begin{eqnarray}
J_u=-J_v=p.
\label{eq.3.16}
\end{eqnarray}
Namely, $J_u$ and $J_v$ are conserved everywhere in the low energy region. When we evaluate $J_u$ and $J_v$ of the incident wave, one finds $J_u=-J_v=p+O(p^2)$ for $x\to -\infty$. This also confirms the perfect transmission of $J_u$ and $J_v$, as well as Bogoliubov phonon. 
\par
We note that, although Eq. (\ref{eq.3.16}) looks different from Eq. (\ref{eq.22}) (where $J_u$ and $J_v$ are constant in the low energy limit), the reason is simply due to the assumed magnitude of the incident wave in Eq. (\ref{eq.3.2}). When we choose the magnitude of the incident wave $(u_I,v_I)$ far away from the barrier so as to be equal to the plane wave solution $(u_p,v_p)$ given by Eq. (\ref{eq.21}), Eq. (\ref{eq.3.16}) is replaced by $J_u=-J_v=1/2$. Under this normalization, a finite quasiparticle current $J_{\rm qp}$ obtained from higher order terms in $J_u$ and $J_v$ with respect to $p$ is proportional to $p$, as in the uniform system discussed in Sec. II. It has been shown\cite{Tsuchiya} that this finite $J_{\rm q}$ is not conserved near the barrier due to the induction of supercurrent counterflow. 
\par

\begin{figure}
\centerline{\includegraphics[width=10cm]{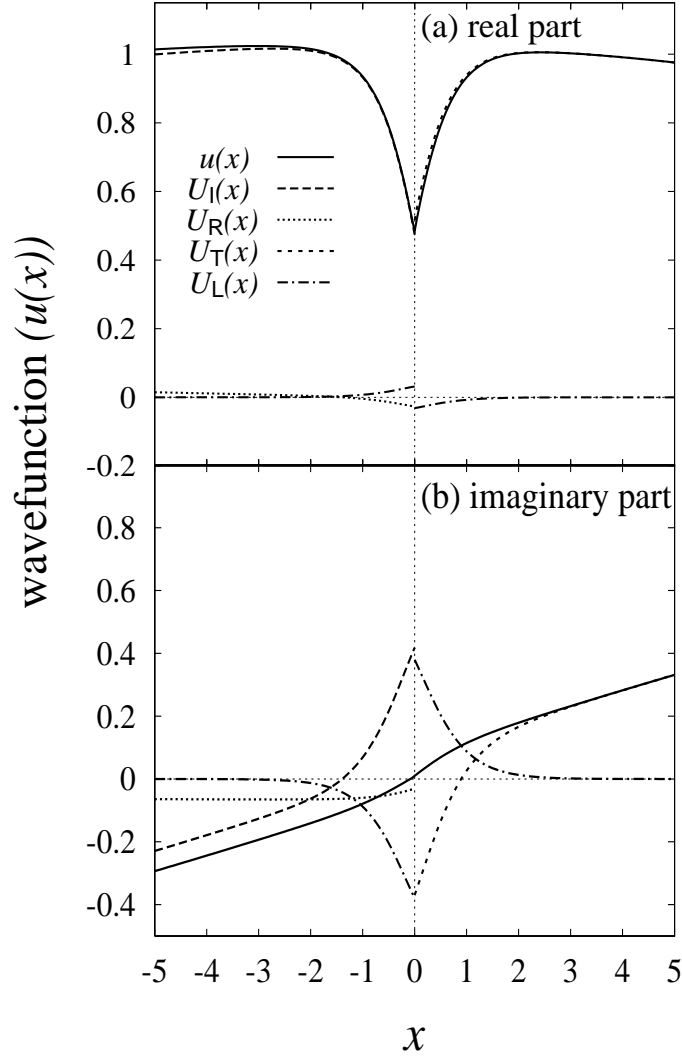}}
\caption{Spatial variation of $u(x)$ given by Eqs. (\ref{eq.3.2}) and (\ref{eq.3.3}). We also show the incident wave component $U_I(x)\equiv u_I(x)e^{ipx}$, reflected wave $U_R(x)=i\alpha p u_R(x)e^{-ipx}$, transmitted wave $U_T(x)=(1-i\eta p)u_T(x)e^{ipx}$, and localized component $U_L(x)= {\rm sgn}(x)(i/2+\beta p)u_L(x)$. We take $p=0.05$ and $x_0=0.5$.
}
\label{fig2}
\end{figure}

\par
Figure \ref{fig2} shows the incident, reflected, transmitted, and localized components of $u(x)$ in Eqs. (\ref{eq.3.2}) and (\ref{eq.3.3}). Although the perfect transmission looks as if the potential barrier is transparent for low-energy Bogoliubov phonon, the reflected component $U_R(x)\equiv i\alpha pu_R(x)e^{-ipx}$, as well as the localized component $U_L(x)\equiv{\rm sgn}(x)(i/2+\beta p)u_L(x)$, actually contribute to the solution. Indeed, these components are necessary to satisfy the boundary conditions at $x=0$. However, since the coefficient of the reflected component $U_R(x)$ is proportional to $p$, the plane wave factor $e^{-ipx}$ in this component is actually irrelevant in the present treatment within $O(p)$. Thus, although Eq. (\ref{eq.3.12}) involves the reflected wave component, the reflected ``plane wave"  is not actually involved in it. 
\par
We note that, while the localized state described by the last terms in Eqs. (\ref{eq.3.2}) and (\ref{eq.3.3}) does not carry current by itself, it still contributes to the tunneling current through the coupling with the propagating wave in the wavefunction. To see this, we divide $J_u$ into contributions coming from each wave component and their couplings. Then we obtain 
\begin{eqnarray}
J_u=J_P+J_{PL}.
\label{eq.3.17}
\end{eqnarray}
Here, $J_P\equiv{\rm Im}[U_P(x)^*\partial_x U_P(x)]$ is the contribution from the right-going wave $U_P(x)\equiv \theta(-x)u_I(x)e^{ipx}+\theta(x)(1-i\eta p)u_T(x)e^{ipx}$, while $J_{PL}\equiv{\rm Im}[U_P(x)^*\partial_x U_L(x)+U_L(x)^*\partial_x U_P(x)]$ describes coupling effects between the right-going wave and the localized state. The other current components involving the reflected wave $U_R(x)$, such as  $J_R\equiv{\rm Im}[U_R(x)^*\partial_x U_R(x)]$,  $J_{RP}\equiv{\rm Im}[U_R(x)^*\partial_x U_P(x)+U_P(x)^*\partial_x U_R(x)]$, and  $J_{RL}\equiv{\rm Im}[U_R(x)^*\partial_x U_L(x)+U_R(x)^*\partial_x U_L(x)]$, can be ignored in the low energy limit. Calculating $J_P$ and $J_{PL}$, we obtain
\begin{eqnarray}
\left\{
\begin{array}{l}
\displaystyle
J_P=p+{1 \over 4}\Bigl(1-\gamma(x)^4\Bigr),\\
\displaystyle
J_{PL}=-{1 \over 4}\Bigl(1-\gamma(x)^4\Bigr).
\end{array}
\right.
\label{eq.3.18}
\end{eqnarray}
\par
\begin{figure}
\centerline{\includegraphics[width=10cm]{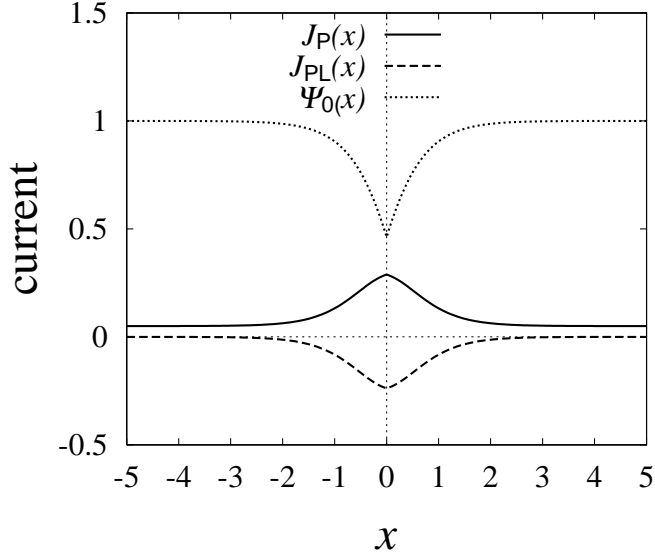}}
\caption{Effects of localized state $U_L(x)$ on $J_u$. We take $p=0.05$ and $x_0=0.5$.
}
\label{fig3}
\end{figure}

Figure \ref{fig3} shows the contribution of $J_P$ and $J_{PL}$ to the current $J_u$. When we ignore effects of localized state $U_L(x)$, the current $J_u$ $(=J_P)$ is not conserved near the potential barrier, where the suppression of the condensate wavefunction $\Psi_{q=0}(x)$ is remarkable. The enhancement of $J_P$ is cancelled out by the counterflow $J_{PL}$ $(<0)$ originating from the coupling of the propagating wave $U_P(x)$ with the localized state $U_L(x)$.

\par
Since $u(x)$ and $v(x)$ have the same form as the condensate wavefunction $\Psi_p(x)$, we can expect that the Bogoliubov equations (\ref{eq.11}) and (\ref{eq.12}) are also related to GP equation (\ref{eq.3}) in the low energy region. To see this, it is convenient to note that the Bogoliubov equations (\ref{eq.11}) and (\ref{eq.12}) can be {\it formally} written as
\begin{equation}
\Bigl(
-{1 \over 2}{d^2 \over dx^2}-(\mu+E)+V\delta(x)+W_u(x)
\Bigr)u(x)=0,
\label{eq.3.19}
\end{equation}
\begin{equation}
\Bigl(
-{1 \over 2}{d^2 \over dx^2}-(\mu-E)+V\delta(x)+W_v(x)
\Bigr)v(x)=0.
\label{eq.3.20}
\end{equation}
Here $W_u$ and $W_v$ are given by
\begin{eqnarray}
\left\{
\begin{array}{l}
\displaystyle
W_u=2|\Psi_q(x)|^2-\Psi_q(x)^2{v(x) \over u(x)},\\
\displaystyle
W_v=2|\Psi_q(x)|^2-\Psi_q^*(x)^2{u(x) \over v(x)}.\\
\end{array}
\right.
\label{eq.3.21}
\end{eqnarray}
In the absence of the supercurrent ($q=0$), setting $\mu=1$, one can rewrite Eqs. (\ref{eq.3.19}) and (\ref{eq.3.20}) in the form
\begin{equation}
\Bigl(
-{1 \over 2}{d^2 \over dy_+^2}-1+{V \over \sqrt{1+p}}\delta(y_+)+{W_u(x) \over 1+p}
\Bigr){\tilde u}(y_+)=0,
\label{eq.3.22}
\end{equation}
\begin{equation}
\Bigl(
-{1 \over 2}{d^2 \over dy_-^2}-1+{V \over \sqrt{1-p}}\delta(y_-)+{W_v(x) \over 1-p}
\Bigr){\tilde v}(y_-)=0.
\label{eq.3.23}
\end{equation}
When $W_u(x)/(1+p)$ and $W_v(x)/(1-p)$ coincide with the nonlinear term in the GP equation, Eqs.(\ref{eq.3.22}) and (\ref{eq.3.23}) reproduce the GP equation (\ref{eq.3}). Indeed, substituting Eqs. (\ref{eq.3.1})-(\ref{eq.3.7}) into Eqs. (\ref{eq.3.21}), we obtain $W_u(x)/(1+p)=|{\tilde u}(y_+)|^2$ and $W_v/(1-p)=|{\tilde v}(y_-)|^2$, within the accuracy of $O(p)$. Thus, as expected, Eqs. (\ref{eq.3.22}) and (\ref{eq.3.23}) have the same form as the GP equation in the supercurrent state with momentum $p$\cite{note2}.
\par
We note that the potential barrier is modified as $V/\sqrt{1\pm p}$ in Eqs. (\ref{eq.3.22}) and (\ref{eq.3.23}), which can explain the reason for the shift of $x_0$ given by $Y_\pm$ in Eqs. (\ref{eq.3.13b}) and (\ref{eq.3.14b}). For example, when one solves the ``GP" equation (\ref{eq.3.22}), the equation for $Y_+$ is obtained as
\begin{equation}
1-\gamma(0,Y_+)^2={V \over \sqrt{1+p}}\gamma(0,Y_+)\simeq 
\Bigl(1-{p \over 2}\Bigr)V\gamma(0,Y_+),
\label{eq.3.24}
\end{equation}
where we have ignored higher order terms than $O(p)$. In Eq. (\ref{eq.3.24}), we are using the notation $\gamma(0,Y_+)$ to emphasize that $x_0$ is different from $Y_+$. Noting that $x_0$ is obtained from the equation $1-\gamma(0,x_0)^2=V\gamma(0,x_0)$, one can solve Eq. (\ref{eq.3.24}) within the accuracy of $O(p)$. The result is
\begin{eqnarray}
\gamma(0,Y_+)
&=&
\gamma(0,x_0)+{1-\gamma(0,x_0)^2 \over 2(1+\gamma(0,x_0)^2)}\gamma(0,x_0)p
\nonumber
\\
&\simeq&
\gamma(0,x_0+{\gamma(0,x_0) \over 2(1+\gamma(0,x_0)^2)}p).
\label{eq.3.24b}
\end{eqnarray}
Here, we have used the identity in Eq. (\ref{eq.3.9}). Equation (\ref{eq.3.24b}) reproduces $Y_+$ in Eq. (\ref{eq.3.13b}). In the same manner, one can confirm that the shift $Y_-$ of $x_0$ in ${\tilde v}(y_-)$ is due to the modified potential barrier $V/\sqrt{1-p}$ in the ``GP" equation (\ref{eq.3.23}).
\par
\section{Quasiparticle tunneling in the supercurrent state}
\par
In this section, we examine effects of supercurrent on the tunneling of Bogoliubov phonon. Assuming that the momentum $q$ carried by condensate, as well as the incident momentum $p$ of the Bogoliubov mode, are very small, we treat them within the first order. We also assume that the supercurrent $J_s$ is much smaller than the critical current. The critical supercurrent state is considered in Sec. V.
\par
Within the accuracy of $O(q)$, we may still ignore $q$ in Eqs. (\ref{eq.7}) and (\ref{eq.9}). Namely, $\gamma(x)$ and $x_0$ are not modified by supercurrent. In addition, we can also set $E=p$ and $(p_1,p_2,p_3,p_4)=(p,-p,2i,-2i)$ under the assumption of small $p$ and $q$. The Bogoliubov wavefunction has the form
\begin{eqnarray}
\left(
\begin{array}{c}
u(x)\\
v(x)
\end{array}
\right)
=
\left(
\begin{array}{l}
e^{i(qx+\theta_q)}{\hat u}_<(x)\\
e^{-i(qx+\theta_q)}{\hat v}_<(x)
\end{array}
\right)\theta(-x)
+
\left(
\begin{array}{l}
e^{i(qx-\theta_q)}{\hat u}_>(x)\\
e^{-i(qx-\theta_q)}{\hat v}_>(x)
\end{array}
\right)\theta(x),
\label{eq.4.1}
\end{eqnarray}
where
\begin{eqnarray}
\left(
\begin{array}{l}
{\hat u}_<(x)\\{\hat v}_<(x)
\end{array}
\right)
=
\left(
\begin{array}{l}
u_I(x)\\v_I(x)
\end{array}
\right)e^{ipx}
+
i\alpha p
\left(
\begin{array}{l}
u_R(x)\\v_R(x)
\end{array}
\right)e^{-ipx}
+
\Bigl({i \over 2}+\beta p-{i \over 2}q\Bigr)\left(
\begin{array}{l}
u_L(x)\\v_L(x)
\end{array}
\right),
\label{eq.4.2}
\end{eqnarray}
\begin{eqnarray}
\left(
\begin{array}{l}
{\hat u}_>(x)\\{\hat v}_>(x)
\end{array}
\right)
=
(1-i\eta p)
\left(
\begin{array}{l}
u_T(x)\\v_T(x)
\end{array}
\right)e^{ipx}
-
\Bigl({i \over 2}+\beta p-{i \over 2}q\Bigr)\left(
\begin{array}{l}
u_L(x)\\v_L(x)
\end{array}
\right).
\label{eq.4.3}
\end{eqnarray}
Here, coefficients $(\alpha$,$\beta$,$\eta)$ and the localized component $(u_L,u_L)$ are not affected by supercurrent, and they are given by Eqs. (\ref{eq.3.4}) and  (\ref{eq.3.6c}), respectively. In contrast, the incident wave $(u_I,v_I)$, reflected wave $(u_R,v_R)$, and transmitted wave ($u_T,v_T$) are modified to be 
\begin{eqnarray}
\left(
\begin{array}{c}
u_I \\v_I
\end{array}
\right)
=
\left(
\begin{array}{l}
\displaystyle
\gamma(x)+{i \over 2}[1-\gamma(x)^2]+{p \over 2}[\gamma(x)+i]
+i{q \over 2}[1+\gamma(x)^2]\\
\displaystyle
\gamma(x)-{i \over 2}[1-\gamma(x)^2]-{p \over 2}[\gamma(x)-i]
-i{q \over 2}[1+\gamma(x)^2]
\end{array}
\right),
\label{eq.4.4}
\end{eqnarray}
\begin{eqnarray}
\left(
\begin{array}{c}
u_R \\v_R
\end{array}
\right)
=
\left(
\begin{array}{l}
\displaystyle
\gamma(x)-{i \over 2}[1-\gamma(x)^2]+{p \over 2}[\gamma(x)-i]
+i{q \over 2}[1+\gamma(x)^2]
\\
\displaystyle
\gamma(x)+{i \over 2}[1-\gamma(x)^2]-{p \over 2}[\gamma(x)+i]
-i{q \over 2}[1+\gamma(x)^2]
\end{array}
\right),
\label{eq.4.5}
\end{eqnarray}
\begin{eqnarray}
\left(
\begin{array}{c}
u_T \\v_T
\end{array}
\right)
=
\left(
\begin{array}{l}
\displaystyle
\gamma(x)-{i \over 2}[1-\gamma(x)^2]+{p \over 2}[\gamma(x)-i]
-i{q \over 2}[1+\gamma(x)^2]
\\
\displaystyle
\gamma(x)+{i \over 2}[1-\gamma(x)^2]-{p \over 2}[\gamma(x)+i]
+i{q \over 2}[1+\gamma(x)^2]
\end{array}
\right).
\label{eq.4.6}
\end{eqnarray}
\par
To see the supercurrent behavior of low-energy Bogoliubov phonon in the supercurrent state, we expand Eq. (\ref{eq.4.1}) in terms of $p$ and $q$ to the first order. In this procedure, we also expand the phase factor $e^{\pm i\theta_q}$ defined in Eq. (\ref{eq.8}) as
\begin{equation}
e^{\pm i\theta_q}=1\mp i{q \over \gamma_0}+O(q^2).
\label{eq.4.7}
\end{equation}
For example, $u(x\ge 0)$ becomes
\begin{eqnarray}
u(x\ge0)
&=&\gamma(x)+i(p+q){\gamma(x)-\gamma_0 \over \gamma_0}+i(p+q)x\gamma(x)
\nonumber
\\
&+&
ip{\gamma_0 \over 1+\gamma_0^2}\gamma(x)
+{p \over 2}{\gamma_0 \over 1+\gamma_0^2}[1-\gamma(x)^2]
+{p \over 2}\gamma(x)+{p \over 2}x[1-\gamma(x)^2].
\label{eq.4.8}
\end{eqnarray}
Comparing Eq. (\ref{eq.4.8}) with Eq. (\ref{eq.3.8}), we find that Eq. (\ref{eq.4.8}) can be written as, within the accuracy of $O(p)$ and $O(q)$,
\begin{equation}
u(x\ge 0)=\sqrt{1+p}e^{ip{\gamma_0 \over 1+\gamma_0^2}}e^{i(p+q)x}
\Bigl[
\gamma(\sqrt{1+p}x+{p\gamma_0 \over 2(1+\gamma_0^2)})
+i(p+q){\gamma(x)-\gamma_0 \over \gamma_0}
\Bigr].
\label{eq.4.9}
\end{equation}
This is essentially the same form as the condensate wavefunction $\Psi_{p+q}(x)$ in (\ref{eq.3.11}). We also reach the same conclusion for $u(\le 0)$. As a result, we obtain
\begin{equation}
{\tilde u}(y_+)={\tilde \Psi}_{p+q}(y_+,Y_+).
\label{eq.4.10}
\end{equation}
Using the relation $v_p(x)=u_{-p}^*(x)$, we also find
\begin{equation}
{\tilde v}(y_-)={\tilde \Psi}_{p-q}(y_-,Y_-).
\label{eq.4.11}
\end{equation}
\par
Equations (\ref{eq.4.10}) and (\ref{eq.4.11}) indicate that tunneling properties of the current $J_u$ and $J_v$ are the same as those of supercurrent in the low energy region. Namely, the Bogoliubov phonon still shows the perfect transmission in the supercurrent state\cite{Danshita}. Equations (\ref{eq.4.10}) and (\ref{eq.4.11}) also show that the supercurrent $J_s=q$ affects $J_u$ and $J_v$ as $J_u=p+q$ and $J_v=-p+q$. This implies that, in the critical supercurrent state, the ``supercurrent" $J_u$ or $J_v$ exceeds the critical current, leading to the breakdown of the perfect transmission. We will confirm this in Sec. V.  
\par
We note that Eqs. (\ref{eq.3.22}) and (\ref{eq.3.23}) are also valid for the supercurrent state. Substituting Eqs. (\ref{eq.4.10}) and (\ref{eq.4.11}) into Eq. (\ref{eq.3.21}), we obtain $W_u/(1+p)=|{\tilde u}(y_+)|^2$ and $W_v/(1-p)=|{\tilde v}(y_-)|^2$. Thus, Eqs. (\ref{eq.3.22}) and (\ref{eq.3.23}) reduce to the GP equation, as expected. 
\par
\section{Breakdown of anomalous tunneling effect in the critical supercurrent state}
\par
In Ref. \cite{Danshita}, the breakdown of the anomalous tunneling is predicted in the critical supercurrent state, when the barrier $V$ is high enough so that the tunneling of supercurrent can be regarded as the Josephson effect. In this section, we also consider the same situation to see if the absence of the anomalous tunneling in the critical supercurrent state can be understood as that the currents $J_u$ or $J_v$ exceeds the critical value.
\par
We briefly summarize the critical supercurrent $J_c~(=q_c)$ in the Josephson coupling regime. To derive the current-phase relation in this regime, it is convenient to write the condensate wavefunction $\Psi_q(x)$ in the form
\begin{equation}
\Psi_q(x)=\sqrt{q^2+\gamma(x)^2}e^{iqx}
e^{i{\rm sgn}(x){\phi(x) \over 2}}.
\label{eq.5.1}
\end{equation}
Here, the phase $\phi(x)$ is defined by 
\begin{equation}
\phi(x)=2
\Bigl(
\tan^{-1}{\gamma(x) \over q}-\tan^{-1}{\gamma_0 \over q}
\Bigr).
\label{eq.5.2}
\end{equation}
Equation (\ref{eq.5.2}) can be rewritten as 
\begin{equation}
q=\gamma_0\tan{\Phi(x) \over 2},
\label{eq.5.3}
\end{equation}
where $\Phi(x)=\phi(x)-2[\tan^{-1}(\gamma(x)/q)-\pi/2]$. When the barrier potential $V$ is very large, $q$ and $\gamma_0$ are very small and one can expand them in terms of $V^{-1}$\cite{Danshita}. Within the accuracy of $O(V^{-1})$, Eq. (\ref{eq.9}) gives 
\begin{equation}
\gamma_0=V(\gamma_0^2+q^2).
\label{eq.5.4}
\end{equation}
From Eqs. (\ref{eq.5.3}) and (\ref{eq.5.4}), one finds
\begin{eqnarray}
\left\{
\begin{array}{l}
q={1 \over 2V}\sin\Phi,\\
\gamma_0={1 \over 2V}(1+\cos\Phi).
\end{array}
\right.
\label{eq.5.5}
\end{eqnarray}
Since the supercurrent is uniform, the $x$ dependence of $\Phi(x)$ in Eq. (\ref{eq.5.5}) is actually irrelevant. When we take the phase $\phi(x)$ at $x=\infty$ ($\equiv\phi_0$), we obtain the well-known Josephson's current-phase relation, $J_s=(1/2V)\sin\phi_0$ within the accuracy of $O(V^{-1})$\cite{Danshita}. Equation (\ref{eq.5.5}) gives the critical current $J_c$ in the Josephson coupling regime as
\begin{equation}
J_c=q_c={1 \over 2V}.
\label{eq.5.6}
\end{equation}
\par
Now we construct the Bogoliubov wavefunction $(u,v)$. In the following discussion, we always assume $p\ll q\sim \gamma_0\sim 1/2V \ll 1$. The Bogoliubov wavefunction, satisfying the boundary conditions, are given by Eqs. (\ref{eq.4.1})-(\ref{eq.4.6}), where the coefficient $\beta$ in Eq. (\ref{eq.4.3}) is replaced by $\beta'$. The coefficients $(\alpha,\beta,\beta',\eta)$ are given by
\begin{eqnarray}
\left\{
\begin{array}{l}
\displaystyle
\alpha={1 \over 2}-(1+q){\gamma_0 \over \gamma_0^2-q^2},\\
\displaystyle
\beta={1 \over 4}-{1-2q \over 2}{\gamma_0 \over \gamma_0^2-q^2},\\
\displaystyle
\beta'={1 \over 4}-{1-q \over 2}{\gamma_0 \over \gamma_0^2-q^2},\\
\displaystyle
\eta={1 \over 2}-(1-4q){\gamma_0 \over \gamma_0^2-q^2}.
\end{array}
\right.
\label{eq.5.7}
\end{eqnarray}
We note that Eq. (\ref{eq.5.5}) gives $\gamma_0=q_c=1/2V$ in the critical supercurrent state. In Sec. IV, we have expanded the Bogoliubov wavefunction $(u,v)$ in terms of $p$ and $q$, where we have implicitly assumed $q/\gamma_0\ll 1$. (See, for example, Eq. (\ref{eq.4.7}).) In the present case ($J_s\lesssim J_c$), although we can still assume $q\ll 1$, we have to treat $\gamma_0$ as the same order as $q$. Namely, the expansion in terms of $q/\gamma_0$, such as Eq. (\ref{eq.4.7}), is not allowed in the present case. In addition, because of $q\simeq \gamma_0$, we have to carefully treat the factor $1/(\gamma_0^2-q^2)$ in Eq. (\ref{eq.5.7}). However, even in this case, we can show that $u$ and $v$ still have the supercurrent properties {\it near the barrier} at $x=0$. 
\par
To show this, we first consider $u(x)$. Within the accuracy of $O(x)$, ${\hat u}(x\ge 0)$ reduces to
\begin{eqnarray}
{\hat u}(x\ge 0)=
\Bigl(
1+ip{\gamma_0 \over \gamma_0^2-q^2}
\Bigr)
\Bigl(
\gamma(x)-iQ_+-2pq{\gamma_0 \over \gamma_0-q^2},
\Bigr).
\label{eq.5.8}
\end{eqnarray}
where $Q_+=p+q$. We also expand the factor $e^{-i\theta_q}$ in Eq. (\ref{eq.4.1}) to $O(p)$, which gives
\begin{eqnarray}
e^{-i\theta_q}
=
{\gamma_0+iq \over \sqrt{\gamma_0^2+q^2}}
=
{\gamma_0+iQ_+ \over \sqrt{\gamma_0^2+Q_+^2}}
\Bigl(
1-ip{\gamma_0 \over \gamma_0^2+q^2}
\Bigr).
\label{eq.5.9}
\end{eqnarray}
Using Eqs. (\ref{eq.5.8}) and (\ref{eq.5.9}), we obtain\cite{note3}
\begin{equation}
u(x\ge 0)={\gamma(0,Z_+)+iQ_+ \over \sqrt{\gamma(0,Z_+)^2+Q_+^2}}e^{iQ_+x}
[\gamma(x,Z_+)-iQ_+],
\label{eq.5.10}
\end{equation}
where $\gamma(x,Z_+)$ equals $\gamma(x)$ where $x_0$ is replaced by 
\begin{equation}
Z_+=x_0-2pq{\gamma(0,Z_+) \over \gamma(0,Z_+)-q^2}.
\label{eq.5.11}
\end{equation}
\par
In the same way, we can rewrite $u(x\le 0)$ as
\begin{equation}
u(x\le 0)={\gamma(0,Z_-)-iQ_+ \over \sqrt{\gamma(0,Z_+)^2+Q_+^2}}e^{iQ_+x}
[\gamma(x,Z_+)+iQ_+].
\label{eq.5.12}
\end{equation}
Comparing Eqs. (\ref{eq.5.10}) and (\ref{eq.5.12}) with Eq. (\ref{eq.6}) we find near the barrier ($x\ll 1$),
\begin{equation}
u(x)={\tilde \Psi}_{p+q}(x,Z_+).
\label{eq.5.13}
\end{equation}
Applying the same analysis to $v(x)$, we obtain
\begin{equation}
v(x)={\tilde \Psi}_{p-q}(x,Z_-),
\label{eq.5.13b}
\end{equation}
where
\begin{equation}
Z_-=x_0+2pq{\gamma(0,Z_-) \over \gamma(0,Z_-)-q^2}.
\label{eq.5.14}
\end{equation}
Thus, we conclude that $u$ and $v$ still have the same properties as those of the condensate wavefunction near the barrier at $x=0$. Their currents are given by $J_u=p+q$ and $J_v=-p+q$ at $x\ll 1$.
\par
Using the fact that $\gamma_0$ satisfies Eq. (\ref{eq.5.4}), we obtains, within the accuracy of $O(p)$
\begin{equation}
\gamma(0,Z_\pm)=V\Bigl(\gamma(0,Z_\pm)^2+Q_\pm^2\Bigr).
\label{eq.5.15}
\end{equation}
We define the phase $\phi_\pm(x)$ ($x\ll 1$) by
\begin{equation}
\phi_\pm(x)=2
\Bigl(
\tan^{-1}{\gamma(x,Z_\pm) \over Q_\pm}
-
\tan^{-1}{\gamma(0,Z_\pm) \over Q_\pm}
\Bigr).
\label{eq.5.16}
\end{equation}
This is analogous to the phase $\phi(x)$ introduced in Eq. (\ref{eq.5.2}). From Eqs. (\ref{eq.5.15}) and (\ref{eq.5.16}), we obtain
\begin{eqnarray}
\left\{
\begin{array}{l}
Q_\pm={1 \over 2V}\sin\Phi_\pm,\\
\gamma(0,Z_\pm)={1 \over 2V}(1+\cos\Phi_\pm),
\end{array}
\right.
\label{eq.5.17}
\end{eqnarray}
where $\Phi_\pm(x)=\phi_\pm(x)-2[\tan^{-1}(\gamma(x,Z_\pm)/Q_\pm)-\pi/2]$. Since the currents $J_u=Q_+=p+q$ and $J_v=-Q_-=-p+q$ are uniform near the barrier, the $x$ dependence of $\Phi_\pm$ in Eq. (\ref{eq.5.17}) can be actually ignored as far as $x\ll 1$. Equation (\ref{eq.5.17}) shows that both $J_u$ and $J_v$ must be smaller than the upper limit of the supercurrent $J_c=1/2V$. However, this condition cannot be satisfied in the critical supercurrent state ($q=J_c=1/2V$). Namely, $J_u$ or $J_v$ always exceeds $J_c$, depending on the direction of the momentum $p$ of Bogoliubov phonon. As a result, the perfect transmission is not obtained in the critical supercurrent state. We emphasize that this gives simple explanation for the breakdown of the anomalous tunneling in the critical supercurrent state predicted in Ref. \cite{Danshita}. 
\par
We note that the coefficients $(\alpha,\beta,\beta',\gamma)$ in Eqs. (\ref{eq.5.7}) diverge when $q=\gamma_0=1/2V$. This means that the components $u$ and $v$ in the Bogoliubov wavefunction no longer have the same form as the condensate wavefunction in the critical supercurrent state. In this case, $u$ and $v$ would have different forms from the condensate wavefunction, leading to a finite reflection probability at the barrier. 
\par
\section{Summary}
In this paper, we have investigated tunneling properties of low-energy Bogoliubov excitations in a Bose superfluid. Using the exactly solvable tunneling problem through a $\delta$-functional potential barrier, we have clarified that, in the low energy region, the components $u(x)$ and $v(x)$ in the Bogoliubov wavefunction $(u(x),v(x))$ have the same form as the condensate wavefunction in the supercurrent state. Based on this result, we have given physical explanation for the anomalous tunneling effect of Bogoliubov phonon, as well as the absence of this phenomenon in the critical supercurrent state, in a consistent manner.
\par
The current $J_u$ and $J_v$ associated with $u(x)$ and $v(x)$, respectively, tunnel through the barrier without reflection, as in the case of supercurrent tunneling. This gives physical explanation for the perfect transmission of the Bogoliubov phonon predicted recently. We also showed that the supercurrent behaviors of $J_u$ and $J_v$, as well as the resulting perfect transmission of Bogoliubov phonon, are also obtained in the supercurrent state unless the magnitude of the supercurrent reaches the critical value. 
\par
In the Josephson coupling regime (which is realized when the tunneling barrier is very high), the upper limit of $J_u$ and $J_v$ is given by the critical current $J_c$ of the ordinary supercurrent. In the critical supercurrent state, $J_u$ or $J_v$ always exceeds $J_c$, irrespective of the direction of the momentum of Bogoliubov phonon. Because of this, the perfect transmission of Bogoliubov phonon is no longer obtained in the critical supercurrent state. This can explain the reason why the breakdown of the anomalous tunneling effect occurs in the critical supercurrent state. 
\par
In this paper, we have used a simple $\delta$-functional potential barrier. While this model enables us to treat the GP and Bogoliubov equations analytically, any realistic potential barrier should have a finite potential width. However, although our model is simple, we still expect that the essence of our results in this paper would be valid for the case with a more realistic barrier potential. We will extend our study to more general and realistic cases in our future paper. Since the Bogoliubov phonon is one of the most fundamental phenomena in the superfluid phase, the observation of the anomalous tunneling would be important in understanding basic properties of this collective mode. 

\begin{acknowledgments} 
We would like to thank I. Danshita for useful discussions on the anomalous tunneling effect in the supercurrent state. This work was supported by a Grant-in-Aid for Scientific research from MEXT (18043005) and CTC program.
\end{acknowledgments}



\begin{thebibliography}{}
\bibitem{note} Far away from the barrier where effects of the barrier potential can be ignored, the supercurrent velocity $v_s=q/m$ must be smaller than the Bogoliubov phonon velocity $v_\phi=\sqrt{gn_0/m}$, because of the well known Landau criterion for the stability of the supercurrent state. When one uses the dimensionless variables used in the text, this inequality is written as $q<1$. (Note that the momentum is scaled by $\xi^{-1}$ in this paper.) Thus, Eq. (\ref{eq.17b}) does not diverge.
\bibitem{note9} In Eq. (\ref{eq.20}), the signs of $J_v$ and $J_u$ are chosen to be opposite to each other so as to emphasize that they are counter-flowing currents. With this regard, we note that the definition of $J_v$ in this paper is different from that in Ref. \cite{Tsuchiya}, where it is defined by $J_v={\rm Im}[v^*(x)\partial_x v(x)]$. The resulting probability current density and quasiparticle density in Ref. \cite{Tsuchiya} are given by $J=J_u+J_v$ and $J_{\rm qp}=J_u-J_v$, respectively. We also note that the probability current density $J$ is related to the energy flux $Q$ discussed in Ref.\cite{Tsuchiya} as $J=EQ$, where $E$ is the energy of Bogoliubov phonon.
\bibitem{note2} In the supercurrent state, the chemical potential is given by $\mu=1+p^2/2$. Thus, as far as we consider the low energy region within the accuracy of $O(p)$, we may set $\mu=1$ even in the supercurrent state.
\bibitem{note3} In Eq. (\ref{eq.5.10}), although the phase factor $Q_+x$ is a higher order term because of $Q_+\ll1$ and $x\ll1 $, we retain it to compare $u(x)$ with the condensate wavefunction. We note that, even when the factor $e^{iQ_+x}$ is ignored, one can still obtain the tunneling current $J_u=Q_+$ at $x=0$ from the derivative of $\gamma(x,Z_+)$ in Eq. (\ref{eq.5.10}).
\end{thebibliography}
\end{document}